\documentclass{ws-p8-50x6-00}

\newcommand{\Tr}{\mathop{\rm Tr}\nolimits}
\def\v0{\vec{\mbox{$0$}}}

\begin{document}

\title{Lattice simulations for the
running coupling constant of QCD}

\author{A. Cucchieri}

\address{IFSC---S\~ao Paulo University, Caixa Postal 369,
13560-970 S\~ao Carlos SP, Brazil\\
E-mail: attilio@if.sc.usp.br}


\maketitle

\abstracts{
The strong coupling constant $\alpha_s(\mu_0)$,
taken at a fixed reference scale $\mu_0$,
is the single free parameter of QCD and should be known to
the highest available precision. The value of $\alpha_s$ should
also be determined with good accuracy
over as large a range of scales as possible, in order to reveal
potential anomalous running in the strength of the strong
interaction. Lattice QCD
is now able to calculate $\alpha_s$ with accuracy
comparable to or better than experiment. We review the status of
such lattice calculations in quenched and full QCD.}

%
%

\section{QCD coupling and $\beta$-function}\label{subsec:qcd}

The scale dependence of the QCD coupling $\alpha_s =
g^2_s / (4 \pi)$ is controlled by the $\beta$-function:
\be
\mu \, \frac{\partial \alpha_s}{\partial \mu} \,=\,
\beta(\alpha_s) \,=\,
-\left( \frac{\beta_0}{2 \pi} \alpha_s^2 \,+\, 
        \frac{\beta_1}{4 \pi^2} \alpha_s^3 \,+\, 
	\frac{\beta_2}{64 \pi^3} \alpha_s^4
\,+\, \ldots \right)
\; \mbox{,}
\ee
where $\beta_0 = 11 - 2 N_f / 3$, $\,\beta_1 = 51 - 19 N_f / 3$
and $\,N_f\,$ is the number of flavors of quarks with mass less
than the energy scale $\,\mu\,$. The coefficients $\beta_n$ for
$n \geq 2$ are renormalization-scheme-dependent.

By solving the differential equation for $\alpha_s$, a constant
of integration is introduced. This constant is the one fundamental
constant of QCD. Usually one chooses for this constant the value
of $\alpha_s$ at a fixed reference scale $\mu_0$.
Equivalently one can consider the dimensional parameter $\Lambda$,
defined for example as
\begin{eqnarray}
\Lambda(N_f) \, & = & \, \mu \,
   \exp{\left[ \frac{- 2 \pi}{\beta_0\, \alpha_s(\mu)} \right]}
\, \left[ \frac{\beta_0\, \alpha_s(\mu)}{2 \pi} 
        \right]^{- \beta_1 / \beta_0^2}
 \nonumber \\[0.2cm]
& & \;\; \times 
     \exp{ \left\{ - \int_0^{\alpha_s(\mu)} \! du 
       \left[ \frac{1}{\beta(u)} \,+\, 
              \frac{2 \pi}{\beta_0 u^2} \,-\,
   \frac{\beta_1}{\beta_0^2 u} \right] \right\}} 
   \;\mbox{.}
   \label{eq:lambda}
\end{eqnarray}
Note that, by using this relation
at the leading order, one gets that
$\,\Lambda(N_f)\,$ is known with an accuracy of
about $\,5\%\,$ if
$\,\alpha_s(\mu)\,$ is given with an accuracy of about $\,1\%$.

When two different renormalization schemes are considered one can 
write
\be
\alpha_s^{(1)}(\mu) \;= \;\alpha_s^{(2)}(\mu) \left\{ 1 \,+\, a 
                     \frac{\alpha_s^{(2)}(\mu)}{4 \pi} \,+\, b 
		  \left[\frac{\alpha_s^{(2)}(\mu)}{4 \pi}\right]^2
		     +\,\ldots \right\}
\label{eq:alfamu}
\ee
and check that
\begin{eqnarray}
\beta_2^{(1)} \, & = & \, \beta_2^{(2)}
                         \,-\,4 a \beta_1
			 \,+\, 2 (b\,-\,a^2) \beta_0
                            \\[0.2cm]
\Lambda^{(1)}(N_f) \, & = &\, \Lambda^{(2)}(N_f) \,
                   \exp{\left( \frac{a}{2 \beta_0} \right)}
\;\mbox{.}
\end{eqnarray}
Using the $\Lambda$ parameter we can also obtain a
parametrization of the $\,\mu\,$ dependence of the QCD
coupling\footnote{Note that this parametrization is not unique
but depends on the choice of the constant of integration (see
for example problem 13.11 in Ref.\ \protect\cite{lebellac}).}
\be
\alpha_s(\mu) \; =\; \frac{2 \pi}{\beta_0 \log{(\mu / \Lambda)}}
\left[ 1 - \frac{\beta_1}{\beta_0^2}
\frac{\log{(\log{(\mu / \Lambda)})}}{
\log{(\mu / \Lambda)}} + \ldots \right]
\;\mbox{.}
\label{eq:awithm}
\ee
Clearly the coupling decreases as the scale $\mu$ increases,
corresponding to the property of
asymptotic freedom. However, due to the logarithmic
dependence one needs to consider a large range of
scales $\mu$ in order to see a variation of $\alpha_s(\mu)$ by
a factor two or three (see Fig.\ 9.2 in Ref.\ \cite{pdg}).

%
%

\subsection{Experimental results for $\alpha_s(\mu)$}
\label{subsec:alphaexp}

The value for a quantity
$\,A\,$ from an experiment can be used to extract
a value of $\alpha_s(\mu)$ by considering the following
perturbative expansion (in some renormalization scheme)
\be
A  \,= \,c_0 \, + \, c_1  \alpha_s(\mu) 
                 \, + \, c_2 \alpha_s^2(\mu)
		 \, + \, \ldots \
		 \;\mbox{.}
		 \label{eq:Aexp}
\ee
If the scale involved is low one should also try to take into
account non-pertubative contributions to $\,A\,$.
Clearly the physical result $\,A\,$ does not depend on the chosen
renormalization scheme.
However, the truncated perturbative series does
exhibit renormalization-scheme dependence. In particular, the finite
coefficients $\,c_i\,(i \geq 2)$ depend
on the renormalization scale
$\,\mu\,$ and, implicitly, on the chosen renormalization scheme.
Therefore, for a given scheme, one should find the best
choice\footnote{See
Ref.\ \protect\cite{pdg2} for various
methods proposed for choosing the scale $\,\mu$.}
for the scale $\,\mu$ in order to reduce the scale
ambiguity in the expansion (\ref{eq:Aexp}).

In order to compare the values of $\alpha_s(\mu)$ from
various experiments\cite{pdg2,review1}, 
these values must be evolved
--- using the renormalization group --- to a common scale (usually
the mass of the $Z^0$ boson) and related
to a common scheme (usually the
$\overline{MS}$ scheme). The present world-wide average\cite{pdg2}
in the $\overline{MS}$ scheme for $\alpha_s(M_Z)^{(N_f=5)}$ is
$0.1172 \pm 0.002$.

%
%

\section{Numerical evaluation of $\alpha_s(\mu = q)$}
\label{sec:numalf}

One can perform the experiment also on a computer, using
Lattice QCD.\cite{creutz}
For example, one can evaluate numerically the plaquette
$ \, W_{1,1}\;=\;<  \Tr U_{\mu\nu} / 3 > \,$ --- which corresponds 
to the elementary gluonic action on the lattice --- 
and compare the result with a perturbative expansion.
Also in this case one should be careful with the choice
of the scheme used for the perturbative expansion and 
with the choice
of the scale $\,\mu$. For example, it is well known\cite{lepage}
that the expansion of the plaquette $\,W_{1,1}\,$ in terms
of the lattice bare coupling $\,\alpha_0 = g_0^2 / 4 \pi\,$
yields a very poor result. In fact,
for $\,g_0^2 = 1\,$ one-loop perturbation theory
with $\,N_f = 0\,$ predicts
$\, - \log{W_{1,1}} \approx 0.4228\,$ while the corresponding 
Monte Carlo result is about $\,0.5214$.

A better expansion\cite{lepage} is obtained if one considers an
effective coupling defined in terms of a physical
quantity. For example, if $V(q)$ is the inter-quark
potential one can define
\be
V(q) \,=\, - \frac{16 \pi \alpha_V(q^2)}{3 q^2}
\ee
and work in the $\alpha_V$ scheme. In this scheme
one gets, at the one-loop level and with $N_f = 0$,
\be
- \log{W_{1,1}} \, = \, c_1\, \alpha_V(q^2) \, \left[ 1 \, +
     \, \tilde{c}_2\, \alpha_V^2(q^2) \, + \, \ldots \right]
\;\mbox{,}
\label{eq:wexp}
\ee
where $\,\tilde{c}_2 =
\frac{11}{4 \pi} 
   \log{\left(\frac{\pi^2}{a^2 q^2}\right)} + \tilde{d} \,$ and
$\,\tilde{d} \approx 1.329\,$. In order to fix the value
of the scale $\,q$ the idea introduced in Ref.\ \cite{lepage} is
to impose the relation
\be
\alpha_V(\bar{q}^2)\, c_1 \,=\, \alpha_V(\bar{q}^2) \int
                            d^4q \;f(q) \,=\,
                     \int
		                 d^4q \;\alpha_V(q^2) \,f(q)
\;\mbox{,}
\ee
namely $\,\bar{q}\,$ is the average momentum entering in the
one-loop calculation.
By expanding both $\,\alpha_V(\bar{q}^2)\,$ and $\,\alpha_V(q^2)\,$
in terms of $\,\alpha_V(\mu^2)\,$ one gets
\be
 \log{(\bar{q}^2)} \, \int  d^4q \; f(q) \,=\,
     \int  d^4q \; \log{(q^2)} \, f(q)
     \;\mbox{.}
\ee
In the case of the plaquette this prescription gives
$\,\bar{q} \approx 3.4018 / a\,$ where $\,a\,$ is the
lattice spacing.
Clearly, this procedure can be repeated for different Wilson
loops $\,W_{r,t}$, i.e.\ from the numerical values for
$\,W_{r,t}\,$ one can obtain $\alpha_V(\bar{q}^2_{r,t})$
using a perturbative expansion like eq.\ (\ref{eq:wexp}).
As can be seen from
Fig.\ 8 in Ref.\ \cite{lepage}, the coupling $\alpha_V(q^2)$
obtained from such calculations
runs with $\,q\,$ in agreement with a two-loop formula
(\ref{eq:awithm}).

In order to compare these results with experimental data one needs
to fix the lattice spacing $\,a$, i.e.\ the
scale $\,\bar{q}\,$ should be given in physical units. This can be
done by comparing lattice data with an experimental input,
as has been done\cite{davies} using the $\Upsilon$ meson spectrum,
known experimentally with very good accuracy and for which accurate
simulations can be done. Using results for
$N_f = 0$ and $N_f = 2$ dynamical fermions, which allows
extrapolation to physical case $N_f = 3$, the authors
of Ref.\ \cite{davies} obtained a
final result --- in the $\overline{MS}$ scheme
and evolved perturbatively
to the $Z^0$ mass scale --- of $0.1174 \pm 0.0024$.
Similar results have been obtained by other groups, using
different lattice setups, and have been included by the 
Particle Data Group\cite{pdg2} in the evaluation of
the world-wide 
average of $\,\alpha_s(M_Z)^{(N_f=5)}$.

%
%

\subsection{Ingredients for a numerical evaluation of
the QCD running coupling constant}\label{subsec:ricetta}

From the previous analysis it is clear that in order to
evaluate the QCD running coupling constant using
lattice numerical simulations one needs:\footnote{See Ref.\
\protect\cite{review2} for a
previous review of the numerical determination of
$\,\alpha_s$.}

\begin{itemize}
\item a good definition for $\alpha_s$
      allowing high accuracy, control over finite-size
      effects and discretization effects,
\item an accurate determination of the energy scale
      (i.e.\ the lattice spacing $\,a\,$),
\item a perturbative relation with $\alpha_{\overline{MS}}$,
\item the possibility of extending the procedure
      to dynamical fermions (unquenched simulations).
\end{itemize}

%
%

\section{Running coupling in a finite box}\label{sec:alphacoll}

The Alpha collaboration\cite{alphasite}
studies the QCD running coupling using
finite boxes (with physical size $\,L\, = N a$),
a definition for $\,\alpha_s\,$
that explicitly depends on the scale $\,L$, and
a non-perturbative finite-size scaling technique.\cite{alphabox}
To this end one could consider, for example,
the force between static quarks separated by a
distance $\,r = L/2$, given by
\be
\alpha_{q\bar{q}}(L) \,=\, \left.
    \frac{3}{4} r^2 F_{q\bar{q}}(r,L)  \,\right|_{r = L/2}
\label{eq:alphaq}
\;\mbox{,}
\ee
or a correlation function\cite{roma} for a separation
$\,r = L/2\,$. Most of the simulations by the Alpha
collaboration have been carried out using as definition for
$\,\alpha_s$ the response of the system to a change of the
boundary conditions in the Schr\"odinger
functional approach (for details see Ref.\ \cite{SF}).
In all cases
the physical size $\,L\,$ of the lattice
is the only physical scale on which these couplings depend,
i.e.\ the box size may be regarded as a reference scale,
similar to the renormalization scale $\,\mu\,$. Thus, one
has a coupling running with $\,\mu = 1/L$.

The key point in considering these couplings is the use of a
non-perturbative renormalization group technique.
More specifically, one
defines the step-scaling function
$\,\sigma(u)\,$
through the relation
$\,\alpha_s(2 L) \,=\, \sigma(\alpha_s(L))$.
Clearly this function provides the value of the coupling in a
box of size $\,2 L\,$ given a value for the same coupling
in a box of size $\,L$. Thus, $\,\sigma(u)\,$ can be regarded
as an ``integrated'' $\beta$-function.
The step-scaling function can be evaluated non-perturbatively
using numerical simulations by considering several
values of $\,L$, as shown in Fig.\ 4 in Ref.\ \cite{capitani}.
Notice that for each data point in the plot
one should carefully check for possible systematic
effects, especially discretization errors. In fact,
each point is
the result of an extrapolation\footnote{See Ref.\ \cite{ps}
for possible problems with this extrapolation.}
to zero lattice spacing
of data obtained from several simulations, done
with the same physical
size $\,L\,$ but different lattice spacing $\,a\,$ and
number of sites per direction $\,N$.

Once the step-scaling function is known,
i.e.\ after fitting the numerical data,
one can evaluate the running
coupling in a very large box $\,L_0 = L_{max}\,$
and then
find the value of the coupling for boxes of sizes
$\,L_i = L_0/2^i$, using a
recursive procedure and a numerical inversion of the
step-scaling function. [In order to keep errors under
control one should avoid using the fit for the
step-scaling function outside the range of the
numerical data, i.e.\ one should avoid
extrapolating the step-scaling function $\,\sigma(u)\,$
to very small values of $\,u$.]
In this way one obtains the values $\,\alpha_s(\mu_i)\,$
for the running coupling at the scales $\mu_i
= 1 / L_i = 2^i / L_{max}$.

Since all the scales $\,\mu_i\,$ are given in
term of $\,L_{max}$, the last step consists in
setting the scale for the largest lattice, i.e.\
finding the value of $\,L_{max}\,$
in physical units by comparing lattice
data with an experimental input.
Thus, the use of a non-perturbative
finite-size scaling technique avoids the requirement
of a non-perturbative evaluation of the scale $\,L_i\,$ 
for very small values of $\,L_i$: this is in general
complicated by finite-size effects and by the fact
that small values of $\,L_i$ correspond to large
value of $\beta$ and that simulations at large $\beta$
and small physical volumes are essentially
``perturbative''.

The final value\cite{capitani} obtained by the Alpha collaboration
in pure QCD is
$\Lambda^{(N_f=0)}_{\overline{MS}} = 251 \pm 21$ MeV.
Preliminary data with dynamical fermions
($N_f = 2$) have been presented last year\cite{alphadym}.

%
%

\section{Numerical determination of $\Lambda$}

Several groups study the running of the QCD coupling
(in some scheme) over a range of energies and then try to fit
the data using
a two- or three-loop expression for $\alpha(q)$ as in
eq.\ (\ref{eq:awithm}),
considering $\Lambda$ as a fitting parameter.
Alternatively, one can evaluate $\Lambda$ as a function of
$\alpha(q)$ and look for a plateau at high scale [note that
in eq.\ (\ref{eq:lambda}) the left-hand side is constant,
for a given value of $N_f$, even though the right-hand
side depends explicitly on the scale $\,\mu$].

%
%

\subsection{Inter-quark potential}\label{sec:pot}

One can evaluate the QCD coupling using the static
potential $V(r)$, which can be evaluated for
several values of $r$
with good accuracy using Wilson loop data
and the relation\cite{creutz}
\be
W(r,t) \,=\, C(r) e^{[- t V(r)]} \,+\, \ldots
\;\mbox{.}
\ee
This formula should in general
be considered in the limit of large $\,t$,
but using\footnote{For a review of this method
see for example Ref.\ \protect\cite{reviewv}.}
smearing and noise-reduction
techniques one can increase the overlap
of the evaluated quantity with the ground state, i.e.\
increase the value of $C(r)$, and obtain a good signal already
for small values of $\,t$.
Then, from $V(r)$ one can obtain the force $F(r)$
[and the running coupling $\alpha_{q\bar{q}}(1/r)$, see eq.\
(\ref{eq:alphaq})]
using a finite-difference approximation for the derivative
$dV(r) / dr$.
In the quenched case\cite{bali} one obtains
$\Lambda^{(N_f=0)}_{\overline{MS}} = 246 \pm 7 \pm 3$
MeV, in excellent agreement with the result from the Alpha
collaboration.
Unquenched simulations (also interesting
to study string breaking\cite{strbr}) are under 
way\cite{vdym}.

%
%

\subsection{Three-gluon vertex}\label{sec:3g}

In this case one evaluates the three-gluon
vertex\cite{parrinello}
in the ${\widetilde{MOM}}$ scheme (in Landau gauge).
For this quantity perturbative results are known to
three loops\cite{3loops} (and the corresponding beta function to
four loops).
In order to obtain good data a careful analysis
of breaking of rotational symmetry is required\cite{3vertbis}.
From quenched simulations\cite{3vert} one obtains the value
$\Lambda^{(N_f=0)}_{\overline{MS}} = 295 \pm 5 \pm 15$
MeV, somewhat larger than the previous results.
Preliminary unquenched
data\cite{3vertdym} have been presented last year
and the authors quote 
$\Lambda^{(N_f=2)}_{\overline{MS}} = 264 \pm 27$
MeV.
A similar study has started recently using the
quark-gluon vertex\cite{skullerud} yielding
$\Lambda^{(N_f=0)}_{\overline{MS}} = 300^{\;+150}_{\;-180}
\pm 55 \pm 30$ MeV, in agreement with the result
obtained using the three-gluon vertex.

%
%

\subsection{Gluon propagator}

In this case\cite{becirevic}
one evaluates the gluon propagator $\,D(q)\,$
(in Landau gauge) and defines
$\, Z_3(q) = q^2 D(q)$.
Then, the gluon data can be fitted
by considering the two coupled differential equations
\begin{eqnarray}
\frac{d \log{Z_3(q)}}{d \log{q^2}} &= &\Gamma(\alpha)
  \,=\,- \left[ \frac{\gamma_0}{4 \pi} \alpha \,+\,
                \frac{\gamma_1}{(4 \pi)^2} \alpha^2 \,+\, \ldots
	 \right] \\
\frac{d \alpha}{d \log{q}} &= &\beta(\alpha)
  \,=\,- \left[ \frac{\beta_0}{2 \pi} \alpha \,+\,
                  \frac{\beta_1}{(2 \pi)^2} \alpha^2 \,+\, \ldots
	 \right]
\;\mbox{,}
\end{eqnarray}
where $\Gamma(\alpha)$ is the anomalous dimension
of the gluon
renormalization constant in the ${MOM}$ scheme.
At the lowest order these equations imply the well-known result
$\, D(q^2) \sim 1/q^2 \, \log^c(q^2 / \Lambda^2)\,$
with $\,c = \frac{\gamma_0}{2 \beta_0} = 13/22$.

A solution for these two equations depends on the values
$\,Z_3(\mu)\,$
and $\,\alpha(\mu)\,$
given at some scale $\,\mu$, and
a direct fit of the data provides
a value for these two constants.
The value $\alpha_s(\mu)$ obtained from the fit
can then be evolved, using the renormalization group,
to the $Z^0$ scale and related to the $\overline{MS}$ scheme,
using the perturbative relation between the two schemes.
The authors\cite{becirevic} quote
$\Lambda^{(N_f=0)}_{\overline{MS}} = 319 \pm 14^{\;+10}_{\;-20}$
MeV as final result in the quenched case, in agreement
with the results quoted in Sec\ \ref{sec:3g}.

%
%

\subsection{Search for power corrections}

Recently there have been  various
attempts of finding power
corrections to the perturbative running of $\,\alpha_s$,
following theoretical predictions\cite{zakharov}.
In particular data from the inter-quark 
potential\cite{oneoverq2pot},
the three-gluon vertex\cite{oneoverq2tg,oneoverq2gl}
and the gluon propagator\cite{oneoverq2gl}
have been reanalyzed. In all cases the authors
found that a nonzero contribution proportional to
$\,1/p^2\,$ improves the fit of the data. In
particular, the result obtained in this way for
the $\,\Lambda\,$ parameter using the three-gluon
vertex\cite{oneoverq2tg}, i.e.\
$\Lambda^{(N_f=0)}_{\overline{MS}} = 237 \pm 3^{\;0}_{\;-10}$
MeV,
is in much better agreement with the results
quoted in Sections \ref{sec:alphacoll} and \ref{sec:pot}
than the original value (see Section \ref{sec:3g}).

%
%

\section{Running couplings at the IFSC-USP}

We now describe the simulations being performed since July 2001
at the Institute of Physics of the University
of S\~ao Paulo, S\~ao Carlos (IFSC--USP).
In connection with a grant from FAPESP (``Projeto Jovem Pesquisador'', 
together with Tereza Mendes), we have set up a dedicated PC cluster
at the IFSC--USP.
The system has 16 nodes with
866 MHz Pentium III CPU and 256 MB RAM memory
(working at 133 MHz), a server with 866 MHz
Pentium III CPU and 512 MB RAM memory 
(working at 133 MHz) and is operating
with Debian Linux. The machines are connected with a
100 Mbps full-duplex network and in total there are
130 GB in {\tt/tmp} directories and 24.8 GB for the
{\tt/home} directory.

We are carrying out simulations considering two
definitions of the running coupling constant.
In the Coulomb gauge\footnote{Work in collaboration with
D.\ Zwanziger, New York University (NYU). Some of our simulations 
are done on a PC cluster at NYU.} we consider\cite{coul}
\be
g^2_{C}(\vp)
\;\equiv\; { {11} \over {12} }
\vp^2 V(\vp) \;\equiv\;
{ {11} \over {12} }
g^{2}_{0}\,\vp^2\,D_{44}(\vp)
\;\mbox{,}
\ee
where $V(\vp)$
is the color-Coulomb potential, $\,D_{44}(\vp)\,$ is the time-time
component (at equal time) of the gluon propagator and
the momenta are three-dimensional.\footnote{Remarkably,
this definition of the running coupling constant
is renormalization-group invariant\protect\cite{coul}.}
Clearly, if $\,V(\vp)\,$ is governed by a 
string tension at large distances,
i.e.\ goes like $\,1 / \vp^{4}\,$ at small momenta,
then we should find
$\,g_{C}^2(\vp) \sim 1 / \vp^2\,$ in the infrared limit.
In Landau gauge\footnote{Work in collaboration with K.\ Langfeld 
and J.\ C.\ R.\ Bloch, T\"ubingen University.}
we consider\cite{vsmekal}
\be
g^2_{L}(p) \;\equiv\; \frac{g^{2}_{0}}{4 \pi}
\,\left[\,p^2\,D(p)\,\right]
\,\left[\,p^2\,G(p)\,\right]^2
\;\mbox{,}
\ee
where $D(p)$ is the transverse gluon propagator and $G(p)$ is the
ghost propagator. 
This expression enters directly in the quark Schwinger-Dyson
equation\cite{bloch}
and can be interpreted as an effective interaction
strength between quarks.\footnote{This is also a 
renormalization-group-invariant
quantity since (in Landau gauge)
$\,\widetilde{Z_1} \,=\,1$.}
In this case one obtains $\,g^2_L (p) \sim p^{-2}\,$ if, for
example, $\,D(p) \sim \mbox{const}\,$
and $\,G(p) \sim p^{-4}\,$ in the infrared limit. On the contrary,
if the gluon propagator goes to $\,0\,$ in the infrared limit and
the ghost propagator blows up not faster than $\,p^{-4}\,$
then $\,g^{2}_{L}(p)\,$ has an
infrared fixed point. The last behavior has also
been obtained\cite{vsmekal} by solving
(approximatively) a coupled set of Schwinger-Dyson equations.

In our project we
are also interested in a numerical verification
of Gribov's {\it confinement} scenarios --- written in terms
of the infra-red bahavior for the propagators considered above ---
for these two gauges.
Theoretical studies\cite{gribov,vanish} predict that
in Landau gauge, there should be a
strong suppression of the (unrenormalized) transverse
gluon propagator $D(p)$ in the
infrared limit and
an enhancement of the ghost propagator $G(p)$ in
the same limit. These results clearly
indicate the absence of a massless gluon from the physical
spectrum and provide an
indication of a long-range effect in the theory that may result in
color confinement.
The confinement scenario is particularly simple
in the minimal Coulomb gauge where the ghost propagator determines
directly the Coulomb interaction\cite{gribov,coul}. In fact, in
this case, confinement of color, i.e.\ the enhancement at
long range of the color-Coulomb potential $V(R)$, is due to the
enhancement of ghost propagator $G(\vp)$ at small momenta.
At the same time, the disappearance of gluons from the
physical spectrum is manifested by the suppression at
$\vp = \v0$ of the propagator $D_{ij}(\vp, p_4)$ of
3-dimensionally transverse would-be physical gluons.

Preliminary results in the pure $SU(2)$ case
for the two running coupling
constants and Gribov's confinement scenarios
have been presented in Refs.\ \cite{alfaifsc}.

%
%

\section*{Acknowledgments}

The author thanks the organizers of Hadron Physics 2002
for the invitation, and Tereza Mendes for her help in writing
this manuscript.
This work has been
supported by FAPESP, Brazil (Project No. 00/05047-5).


\begin{thebibliography}{99}

\bibitem{lebellac} {\em Quantum and statistical field theory},
                   M.\ Le Bellac (Oxford University Press, Oxford,
		   1995).

\bibitem{pdg} {\tt http://pdg.lbl.gov/2001/figures.html}.

\bibitem{pdg2} Review on ``Quantum chromodynamics'' in
               {\tt http://pdg.lbl.gov/}.

\bibitem{review1} P.\ N.\ Burrows et al., hep-ex/{\tt 9612012}.

\bibitem{creutz} See for example {\em Quarks, gluons and
                 lattices}, M.\ Creutz (Cambridge
		 University Press, Cambridge, 1983).

\bibitem{lepage} G.\ P.\ Lepage and P.\ B.\ Mackenzie,
                 \PRD 48 (1993) 2250.

\bibitem{davies} C.\ T.\ H.\ Davies et al., \PRD 56 (1997) 2755.

\bibitem{review2} P.\ Weisz, \NPB (Proc. Suppl.) 47 (1996) 71.

\bibitem{alphasite} See {\tt 
          http://www.ifh.de/computing/projects/parallel/ape/alpha/
          alpha.html}

\bibitem{alphabox} M.\ L\"uscher, P.\ Weisz and U.\ Wolff,
                   \NPB 358 (1991) 221. Also, see for example
		   R.\ Sommer, hep-ph/{\tt 9711243}, in
                   Schladming 1997 ``Computing particle properties'',
		   and M.\ L\"uscher, hep-lat/{\tt 9802029}.

\bibitem{roma} G.\ M.\ de Divitiis et al., 
	       \NPB 422 (1994) 382; \NPB 433 (1995) 390.

\bibitem{SF} M.\ L\"uscher et al., \NPB 389 (1993) 247.

\bibitem{capitani} S.\ Capitani et al., \NPB (Proc. Suppl.) 63 (1998) 153.

\bibitem{ps} A.\ Patrascioiu and E.\ Seiler,
             hep-lat/{\tt 0009005}.

\bibitem{alphadym} A.\ Bode et al., 
		   \PLB 515 (2001) 49.

\bibitem{reviewv} G.\ S.\ Bali,
                  {\em Phys.\ Rept.} 343 (2001) 1.

\bibitem{bali} G.\ S.\ Bali and K.\ Schilling, \PRD 47 (1993) 661.

\bibitem{strbr} B.\ Bolder et al., \PRD (2001) 63;
                C.\ W.\ Bernard et al., \PRD 64 (2001) 074509.

\bibitem{vdym} G.\ S.\ Bali et al.,
               \PRD 62 (2000) 054503.

\bibitem{parrinello} C.\ Parrinello,
                     \PRD 50 (1994) 4247.

\bibitem{3loops} K.\ G.\ Chetyrkin  and A.\ Retey,
                 hep-ph/{\tt 0007088}. 

\bibitem{3vertbis} P.\ Boucaud et al., JHEP 9810 (1998) 017.

\bibitem{3vert} P.\ Boucaud et al., JHEP 9812 (1998) 004.

\bibitem{3vertdym} P.\ Boucaud et al., JHEP 0201 (2002) 046.

\bibitem{skullerud} J.\ Skullerud, A.\ Kizilersu and A.\ G.\ Williams, 
                    \NPB (Proc.Suppl) 106 \& 107 (2002) 841;
		    J.\ Skullerud and A.\ Kizilersu,
		    hep-ph/{\tt 0205318}. 

\bibitem{becirevic} D.\ Becirevic et al., 
                    \PRD 60 (1999) 094509;
		    \PRD 61 (2000) 114508.

\bibitem{zakharov} V.\ I.\ Zakharov, \NPB (Proc.\ Suppl.) 74
                   (1999) 392.

\bibitem{oneoverq2pot} G.\ S.\ Bali, \PLB 460 (1999) 170.

\bibitem{oneoverq2tg} P.\ Boucaud et al., JHEP 0004 (2000) 006.
                    
\bibitem{oneoverq2gl} D.\ Becirevic et al.,
                      \NPB (Proc.\ Suppl.) 106 \& 107 (2002) 867.

\bibitem{coul} D.\ Zwanziger, \NPB 518 (1998) 237.

\bibitem{vsmekal} L.\ von Smekal, R.\ Alkofer and
                  A.\ Hauck, {\em Phys. Rev. Lett.} 79 (1997) 3591.

\bibitem{bloch} J.\ C.\ R.\ Bloch,
                hep-ph/{\tt 0202073}.

\bibitem{gribov} V.\ N.\ Gribov, \NPB 139 (1978) 1.

\bibitem{vanish} D.\ Zwanziger, \NPB 364 (1991) 127;
                 \NPB 412 (1994) 657.

\bibitem{alfaifsc} A.\ Cucchieri and D.\ Zwanziger, \PRD
                   65 (2002) 014001;
                   A.\ Cucchieri, T.\ Mendes and D.\ Zwanziger,
                   \NPB (Proc. Suppl.) 106 \& 107 (2002) 697;
		   talks to be presented at Lattice 2002 by
                   D.\ Zwanziger ({\em Gluon propagator in 
                   minimal Coulomb gauge}) and A. Cucchieri
	           ({\em Running coupling constant and propagators in
		   $SU(2)$ Landau gauge}).

\end{thebibliography}
\end{document}